\documentclass[12pt]{article}
\usepackage{amsmath,amssymb,epsfig,setspace}
\begin{document}
\title{The Peculiar Phase Structure of Random Graph Bisection}
\author{Allon G.\ Percus\thanks{Information Sciences Group,
Computer, Computational and Statistical Sciences Division,
Los Alamos National Laboratory,
Los Alamos, NM 87545, USA
and Department of Mathematics, UCLA,
Los Angeles, CA 90095, USA,
e-mail: percus@ipam.ucla.edu}
\and
Gabriel Istrate\thanks{eAustria Research Institute,
Bd.\ V.\ P\^arvan 4, cam.\ 045B,
RO-300223 Timi\c{s}oara, Romania,
e-mail: gabrielistrate@acm.org}
\and
Bruno Gon\c{c}alves\thanks{Department of Physics,
Emory University,
Atlanta, GA 30322--2430, USA,
e-mail: bgoncalves@physics.emory.edu}
\and
Robert Z.\ Sumi\thanks{Department of Physics,
Babes-Bolyai University, RO-400884 Cluj, Romania
and eAustria Research Institute,
Bd.\ V.\ P\^arvan 4, cam.\ 045B,
RO-300223 Timi\c{s}oara, Romania,
rsumi@phys.ubbcluj.ro}
\and
Stefan Boettcher\thanks{Department of Physics,
Emory University,
Atlanta, GA 30322--2430, USA,
e-mail: stb@physics.emory.edu}
}

\date{}
\maketitle

\begin{abstract}
The mincut graph bisection problem involves partitioning the $n$
vertices of a graph into disjoint subsets, each containing exactly $n/2$
vertices, while minimizing the number of ``cut'' edges with an endpoint
in each subset.  When considered over sparse random graphs, the phase
structure of the graph bisection problem displays certain familiar
properties, but also some surprises.  It is known that when the mean
degree is below the critical value of $2\log 2$, the cutsize is zero with
high probability.  We study how the minimum cutsize increases with mean
degree above this critical threshold, finding a new analytical upper
bound that improves considerably upon previous bounds. Combined with
recent results on expander graphs, our bound suggests the unusual
scenario that random graph bisection is replica symmetric up to and
beyond the critical threshold, with a replica symmetry breaking
transition possibly taking place \emph{above} the threshold.  An
intriguing algorithmic consequence is that although the problem is
NP-hard, we can find near-optimal cutsizes (whose ratio to the
optimal value approaches 1 asymptotically) in polynomial time
for typical instances near the phase transition.
\end{abstract}

\section{Introduction}

The graph bisection problem arises in a wide range of contexts,
ranging from VLSI design~\cite{alpert-1995-vlsi} to computer
vision~\cite{boykov-2001-ieee,kolmogorov-2004-ieee} to protecting
networks from attack~\cite{holme-2002-attack}.  The basic mincut graph
bisection (or graph bipartitioning) problem is defined as follows.  Given
an undirected, unweighted graph, partition its $n$ nodes into two
disjoint subsets of equal size, while
minimizing the number $w$ of ``cut'' edges with an
endpoint in each subset.  This is an NP-hard optimization
problem~\cite{garey-1979-computers}.

From a statistical physics
perspective, graph bisection is also equivalent to finding
the ground state of a ferromagnet constrained to have zero
magnetization~\cite{fu-1986-application,kanter-1987-meanfield,
mezard-1987-meanfield,liao-1987-prl,liao-1988-pra}.
Over the past decade, the study of combinatorial optimization problems
within the physics community as well as the study of statistical physics
approaches within the computer science community has led to remarkable
successes~\cite{percus-2006-book}.  By considering optimization problems
over appropriately parametrized
ensembles of random instances, physicists have developed methods
leading both to analytical predictions of global optima and to
algorithmic approaches for finding these.  One particularly rich source
of insight has been the investigation of phase transition behavior: for
many combinatorial optimization problems, both the nature of the solution
space and the average-case complexity of algorithms solving the
problem are closely related to an underlying phase structure.

In this paper, we consider graph bisection over Erd\H{o}s-R\'enyi
(Bernoulli) random graphs~\cite{erdos-1959-random} generated from the
${\cal G}_{np}$
ensemble: given $n$ nodes, place an edge between each
of the ${n \choose 2}$ pairs of nodes independently with probability
$p$.  We are interested in the regime of sparse random graphs, where $p$
scales as $1/n$ and so a vertex's mean degree $p(n-1)$ is some finite
$\alpha$.  On such graphs, it is known that a phase transition occurs at
$\alpha_c=2\log
2$~\cite{goldberg-1985-bounds,mezard-1987-meanfield,liao-1987-prl,liao-1988-pra,luczak-2001-bisecting}.
For $\alpha<\alpha_c$, typical graphs can be bisected without
cutting any edges: the {\em bisection width\/} $w$ is zero.
For $\alpha>\alpha_c$, the graph typically contains a connected component
(giant component) of size greater than $n/2$ and so a bisection requires
cutting edges, giving $w>0$.  In other combinatorial optimization
problems such as graph coloring or satisfiability, a similar sharp
threshold coincides with a rapid escalation of average-case computational
complexity~\cite{cheeseman-1991-ijcai,mitchell-1992-aaai,percus-2006-book}.

Additionally, for many of these other problems, at some $\alpha_d\leq
\alpha_c$ the solution space undergoes a structural transition that
corresponds to replica symmetry
breaking~\cite{biroli-2000-variational,mezard-2002-pre}.  Define two
optimal solutions to be
{\em adjacent\/} if the Hamming distance between them is small: an
infinitesimal fraction of the variables differ in value.  Then below $\alpha_d$, any two
solutions are connected by a path of adjacent solutions.  The
solution space may be thought of as a single ``cluster''.  This is a
replica symmetric (RS) phase.  But above $\alpha_d$, the cluster
fragments into multiple non-adjacent clusters, differing by an extensive
number of variables.  This is a replica symmetry breaking (RSB) phase.
In cases where $\alpha_d<\alpha_c$, such as 3-COL and 3-SAT, this
structural transition may contribute to the rise in algorithmic
complexity as the main threshold $\alpha_c$ is approached from
below~\cite{mezard-2002-science}.

We find that the scenario is quite different for graph bisection.
Rather than the RSB transition taking place below $\alpha_c$, our
arguments suggest that it actually occurs somewhere above $\alpha_c$.
Graph bisection is the first combinatorial optimization problem we are
aware of that displays this behavior.  Furthermore, in 1987, Liao
proposed an analytical solution for bisection
width~\cite{liao-1987-prl,liao-1988-pra}
that assumes replica symmetry above $\alpha_c$, so our results help
validate this prediction.  Finally, the same arguments that indicate
replica symmetry also suggest an algorithm for finding extremely close
approximations to the optimal solution, for
$\alpha\in(\alpha_c,\alpha_d)$, in polynomial time.

The rest of the paper is organized as follows.  In
Section~\ref{sec:width} we study the bisection width (or ``cutsize'')
$w$ as a function of mean degree $\alpha$.   We present numerical results and
derive a new upper bound by analyzing a simple greedy heuristic.  In
Section~\ref{sec:clustering} we study the solution cluster structure,
and argue that the problem is replica symmetric through and beyond the
critical threshold $\alpha_c=2\log 2$.  In Section~\ref{sec:discussion}
we conclude with a discussion of the physical and algorithmic
consequences of this unusual phase structure.

\section{Bisection Width}
\label{sec:width}

Formally, the problem is as follows.  Given an undirected, unweighted
graph $G=(V,E)$ where $|V|=n$ is even, a valid solution is a partition
of the vertex set into two disjoint subsets $V_1$ and $V_2$ such that
$|V_1|=|V_2|=n/2$.  An optimal solution is one that minimizes the
{\em bisection width\/} $w=|E(V_1,V_2)|$,
i.e., the number of ``cut'' edges with an endpoint in each subset.
We consider graphs $G$ drawn uniformly at random from ${\cal G}_{np}$,
and take our control parameter to be the graph's mean degree
$\alpha=p(n-1)$.

The ${\cal G}_{np}$ ensemble~\cite{erdos-1959-random,bollobas-2001-book}
has been studied very extensively since its introduction by Erd\H{o}s
and R\'enyi in 1959, and many of its structural properties are known.
These have a crucial effect on the bisection width.
\begin{itemize}
\item  For $\alpha<1$, the
graph consists only of small components, the largest being
asymptotically of size
$O(\log n)$.  The fraction of monomers, or isolated vertices, is
asymptotically almost surely (a.a.s.)\ itself at
least $e^{-1}$.  Consequently, with high probability a perfect bisection
($w=0$) exists.
\item For $\alpha>1$, there is a unique giant component containing
$O(n)$ vertices.  The second-largest component is asymptotically of size
$O(\log n)$.
As long as $\alpha$ is below the threshold $\alpha_c=2\log 2$,
the giant component a.a.s.\ contains a fraction $g<1/2$ of the vertices, and the
fraction
of monomers is still above $1/4$.  This allows one to
prove~\cite{goldberg-1985-bounds,luczak-2001-bisecting} that with high
probability, $w=0$ for all $\alpha<2\log 2$.
\item For $\alpha>2\log 2$, $g>1/2$,
and so bisecting the graph requires cutting some of its
edges.  It is known~\cite{luczak-2001-bisecting} that $w$ is extensive
(scales as $n$) in this regime, as long as $\alpha$ is finite.
\end{itemize}

\begin{figure}
\begin{center}
\includegraphics[height=3in]{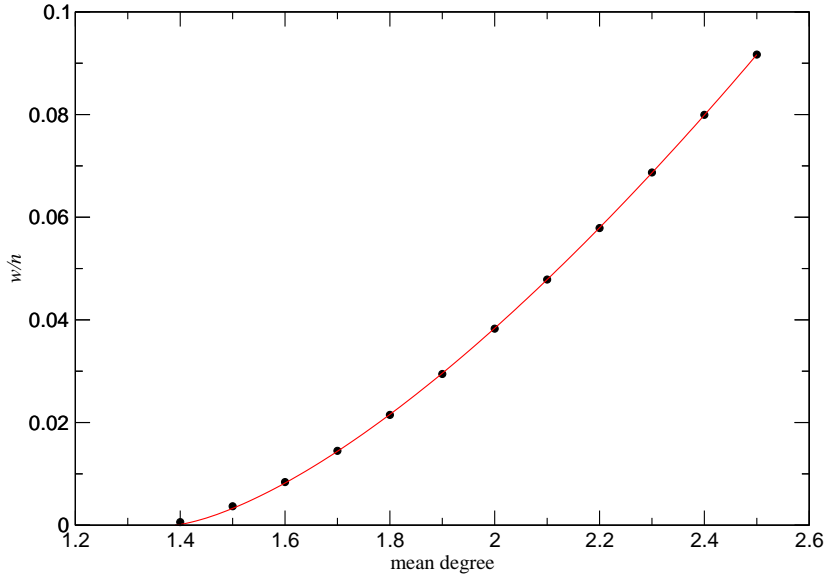}
\end{center}
\vspace{-0.2in}
\caption{Presumed optimal bisection width, found by extremal
optimization on random graphs for $n=2000$.  Line shows fit to
$\gamma(\alpha-\alpha_c)^\beta$, with $\gamma\approx 0.0783$ and
$\beta\approx 1.46$.  See
also~\cite{boettcher-1999-extremal}.\label{fig:eogbp}}
\end{figure}

How does the bisection width constant $w/n$ behave for
$\alpha>\alpha_c$?
Figure~\ref{fig:eogbp} shows simulation results using the
extremal optimization~\cite{boettcher-2000-ai} heuristic to find the presumed
optimal bisection.  These suggest that the bisection width
undergoes a continuous transition at $\alpha_c$, and that its first
derivative is continuous as well.
However, neither of these properties has yet been shown analytically.
We will do so now, by deriving a straightforward upper bound on $w$.
We give the main elements of the derivation here; a more rigorous
mathematical treatment will be provided elsewhere~\cite{istrate-2009-rsa}.

We obtain the upper bound by ``stripping''
trees off of the giant component.  For random graphs with mean degree
$\alpha$, it is known that the fraction $g$ of nodes that are in the
giant component is given a.a.s.\ by~\cite{erdos-1959-random}
\begin{equation}
\label{eq:gc}
g=1-e^{-\alpha g}
\end{equation}
and the fraction $b$ of nodes that are in trees within the giant component (the
{\em mantle\/}~\cite{pittel-1990-tree}) is a.a.s.
\begin{equation}
\label{eq:mantle}
b=\alpha g (1-g).
\end{equation}

\begin{figure}
\begin{center}
\includegraphics[height=3in]{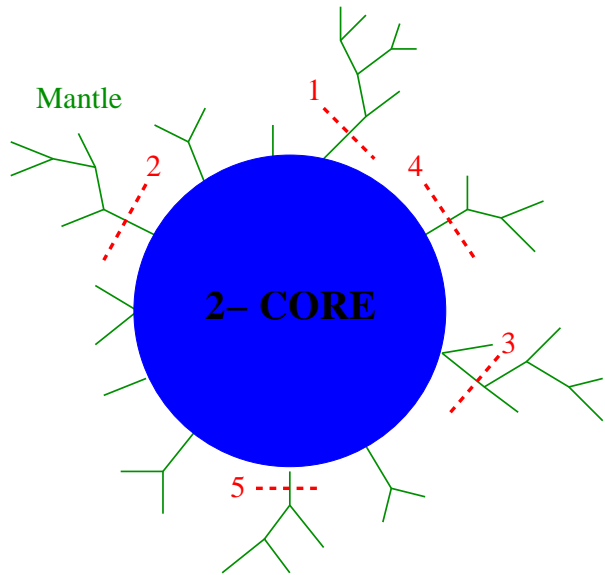}
\end{center}
\vspace{-0.2in}
\caption{Schematic of giant component consisting of a {\em mantle\/} of
trees and a
remaining {\em 2-core\/}.  Greedy upper bound algorithm cuts trees of
decreasing size.  Numbers show order of cuts, depicted by dashed
lines.\label{fig:strip}}
\end{figure}

Now imagine a greedy process (Figure~\ref{fig:strip}) of removing trees,
starting from the largest one and decreasing in size, until the giant
component has been pruned to size $\leq n/2$.
In order to know how many trees need to be removed, consider the
distribution of tree sizes, $P(t)$.  A fortunate consequence of the
probabilistic independence in ${\cal G}_{np}$ is that $P(t)$ is simply
given by the number of ways of constructing a tree of size $t$ from $q$
roots and $r$ other nodes, where $q=(g-b+\sigma(n))n$ is the number of nodes
in the {\em 2-core\/} of the giant component (all nodes {\em not\/} in
the mantle) and $r=(b+\tau(n))n$ is the number of nodes in the
mantle, with $\sigma(n)$ and $\tau(n)$ being random functions that go
a.a.s.\ to zero~\cite{jansson-2000-random,benjamini-2006-mixing}.  A basic
combinatorial argument then gives
\begin{equation}
P(t) = {r \choose t} t^t \frac{q}{r}
\frac{(q+r-t)^{r-t+1}}{(q+r)^{r-1}},
\end{equation}
and for large $n$, letting $\rho=b/g$,
\begin{equation}
\label{eq:treedist}
P(t) \approx \frac{(\rho t)^t e^{-\rho t}}{t!}\frac{1-\rho}{\rho}
\end{equation}
for $t\geq 1$.

Let $z$ be the total number of trees.  Then, given $P(t)$, the total
number of nodes in all
trees is $\sum_{t=1}^\infty t P(t) z$.  As this is simply the size of
the mantle, it holds a.a.s.\ that
\begin{equation}
\label{eq:numtrees}
\frac{z}{n}=\frac{b}{\sum_{t=1}^\infty t P(t)}.
\end{equation}
Let us evaluate this expression.  From normalization of $P(t)$,
\begin{equation}
\sum_{t=1}^\infty \frac{(\rho t)^t e^{-\rho t}}{t!} = \frac{\rho}{1-\rho},
\end{equation}
and differentiating with respect to $\rho$,
\begin{equation}
\left(\frac{1}{\rho}-1\right) \sum_{t=1}^\infty \frac{(\rho t)^t
e^{-\rho t}}{(t-1)!} =
\frac{1}{(1-\rho)^2}.
\end{equation}
The left-hand expression is simply $\sum_{t=1}^\infty t P(t)$, so
Eq.~(\ref{eq:numtrees}) becomes
\begin{equation}
\frac{z}{n}=b(1-\rho)^2.
\end{equation}

We now wish to set $t_0$ so that the number of nodes living on trees of size
$\geq t_0$ is just barely enough to cover the ``excess'' of the giant
component, $\delta=g-1/2$.  In that case, $t_0$ satisfies the property
\begin{eqnarray}
\delta &>&  \frac{z}{n} \sum_{t=t_0+1}^\infty t P(t) \\
\label{eq:deltabound}
&=& \frac{z}{n}\sum_{t=t_0+1}^\infty \frac{(\rho t)^t e^{-\rho t}}{(t-1)!}
\frac{1-\rho}{\rho} \\
&=& \rho e^{-\rho}\frac{z}{n}\sum_{t=t_0}^\infty \frac{(t+1)^{t+1}}{t^t}
\frac{(\rho t)^t e^{-\rho t}}{t!} \frac{1-\rho}{\rho} \\
\label{eq:tzero}
&>& \rho e^{-\rho}\frac{z}{n}\sum_{t=t_0}^\infty 2(t_0+1) P(t)
\end{eqnarray}

Given $t_0$ satisfying both this and the corresponding lower bound on
$\delta$,
bisecting the graph is simply a matter of cutting each tree of size
$\geq t_0$.  Any resulting imbalance in the partitions (from
``overstripping'') can easily be corrected by transferring enough
monomers: there is an extensive number of these, whereas the largest
trees are only of size $O(\log n)$.
Thus,
\begin{eqnarray}
\frac{w}{n} &\leq& \frac{z}{n}\sum_{t=t_0}^\infty P(t) \\
\label{eq:bound0}
&<& \frac{\delta(\rho e^{-\rho})^{-1}}{2(t_0+1)}
\end{eqnarray}
from Eq.~(\ref{eq:tzero}).  In order for the tree-stripping procedure to work,
the mantle must contain
enough vertices to strip, and so we know
that $b\geq\delta$.  One can solve Eqs.~(\ref{eq:gc}) and (\ref{eq:mantle})
numerically to find that this imposes the condition
$g< c_0$ with constant $c_0\approx 0.813$, corresponding
to $\alpha\lesssim 2.06$.
For $\alpha_c \leq \alpha
< 2.06$, it is straightforward to confirm that $\delta(\rho
e^{-\rho})^{-1} < \alpha-\alpha_c$, giving
\begin{equation}
\label{eq:bound1}
\frac{w}{n} < \frac{\alpha-\alpha_c}{2(t_0+1)}.
\end{equation}

Finally, we bound $t_0$ itself.  A version of Stirling's formula gives
the inequality $t!<t^t e^{-t}\sqrt{2\pi t}/(1-1/2t)$, which we apply in
Eq.~(\ref{eq:deltabound}):
\begin{eqnarray}
\delta &>& \frac{z}{n}\sum_{t=t_0+1}^\infty \frac{t (\rho t)^t e^{-\rho t}}{t!}
\frac{1-\rho}{\rho} \\
       &>& \frac{b(1-\rho)^3}{\sqrt{2\pi}\rho}
           \sum_{t=t_0+1}^\infty \frac{\sqrt{t}(1-1/2t)\rho^t e^{-\rho t}}
                                      {e^{-t}} \\
       &>& \frac{b(1-\rho)^3}{\sqrt{2\pi}\rho}
           \sum_{t=t_0+1}^\infty (\rho e^{1-\rho})^t \\
       &=& \frac{b(1-\rho)^3}{\sqrt{2\pi}\rho}
           \frac{(\rho e^{1-\rho})^{t_0+1}}{1-\rho e^{1-\rho}}.
\end{eqnarray}
We have just seen that $\alpha-\alpha_c>\delta(\rho e^{-\rho})^{-1}$
in the regime of interest, so
\begin{equation}
\alpha-\alpha_c > \frac{b(1-\rho)^3}{\sqrt{2\pi}\rho^2
e^{-\rho} (1-\rho e^{1-\rho})} (\rho e^{1-\rho})^{t_0+1}.
\end{equation}
Then letting $\epsilon=\alpha-\alpha_c$,
\begin{equation}
t_0+1 > \frac{\log(1/\epsilon)+
\log\frac{b(1-\rho)^3}{\sqrt{2\pi}\rho^2
e^{-\rho} (1-\rho e^{1-\rho})}}{\log(1/\rho e^{1-\rho})}.
\end{equation}
It can be verified that the second term in the numerator is
monotone increasing in $g$ and is thus bounded below by its value at
$g=1/2$ (the smallest possible giant component size for
$\alpha\geq\alpha_c$), which corresponds to $\rho=\log 2$.  This gives
\begin{equation}
\label{eq:bound2}
t_0+1 > \frac{\log (1/\epsilon)-c_1}{\log(1/\rho e^{1-\rho})}
\end{equation}
where
\begin{equation}
c_1 = -\log\frac{(1-\log 2)^3}{\sqrt{2\pi}\log 2(1-e \log 2/2)} \approx
1.25.
\end{equation}
Finally, the denominator in Eq.~(\ref{eq:bound2}) is also monotone
increasing in $g$, and thus bounded above by its value where the mantle
is exhausted at $g=c_0\approx 0.813$, which corresponds to
$\rho=1-1/2c_0$.  Therefore,
\begin{equation}
\label{eq:bound3}
t_0+1 > \frac{\log (1/\epsilon)-c_1}{c_2}
\end{equation}
where
\begin{equation}
c_2=-\frac{1}{2c_0}-\log \Bigl(1-\frac{1}{2c_0}\Bigr)\approx 0.339.
\end{equation}
Inserting this $t_0$ bound into Eq.~(\ref{eq:bound1}) gives
the bisection width bound
\begin{equation}
\label{eq:bound4}
\frac{w}{n} < \frac{c_2}{2} \frac{\epsilon}{\log (1/\epsilon)-c_1}.
\end{equation}

We stress several points regarding this bound.  First of all, the result
closes the gap between upper and lower bounds on $w/n$ at
$\alpha=\alpha_c$.  The unsurprising fact that both bounds go to zero
and the bisection width is continuous at the transition is already
apparent from Eq.~(\ref{eq:bound0}).  The less obvious property that its
first derivative (with respect to $\alpha$) is also continuous at the
transition follows from Eq.~(\ref{eq:bound4}).  Second of all, 
the denominator of Eq.~(\ref{eq:bound4}) is only positive for
$0 < \epsilon < 1/e^{c_1}$.  Numerically, this is roughly $1.39 < \alpha
< 1.67$, which covers a significant fraction of the regime of interest
$1.39 < \alpha < 2.06$.  Finally, from Eq.~(\ref{eq:bound3}) it is clear
that $t_0$ diverges at the transition, and for sufficiently small $\epsilon$
it satisfies the even simpler bound $t_0>\log(1/\epsilon)$. 
These facts will be significant for the arguments in the next
section.

\section{Solution Cluster Structure}
\label{sec:clustering}

We now consider the geometry of the solution space for random graph
bisection.  For a graph $G\in{\cal G}_{np}$, given two distinct optimal
bisections $A$ and $B$, define the relative Hamming distance $\mu(A,B)$
to be the fraction of nodes that are in one of the subsets ($V_1$ or
$V_2$) in solution $A$ but in the opposite subset in solution $B$.
Since there is a global
symmetry under exchange of $V_1$ and $V_2$, we are interested in
$d(A,B)=\min(\mu(A,B),1-\mu(A,B))$.

Define $A$ and $B$ to be $r${\em-adjacent\/} if $d(A,B)\leq r$.  For
some small $r$, any two solutions that are connected by a chain of
$r$-adjacent solutions are said to be in the same {\em solution
cluster\/}.  Asymptotically, for a given mean degree $\alpha$, we would
like to know what the limiting cluster structure is when $r$ becomes an
arbitrarily small constant.  One scenario that is common for
sufficiently underconstrained combinatorial optimization problems is
{\em replica symmetry\/}: for any finite $r>0$, all solutions are with high
probability within a single cluster.  Another scenario, common for more
constrained instances, is {\em one-step replica symmetry breaking\/}: for
sufficiently small $r$, there is a large number of disconnected
clusters, separated from each other by at least some finite relative
Hamming distance.

In what follows, if solutions constructed in a particular way are with
high probability $r$-adjacent for any finite $r>0$, we will refer to them
as being simply {\em adjacent\/}.

\subsection{$\alpha<2\log 2$}
\label{sec:alphasmall}

Consider graphs where optimal bisections
require no cuts in the giant component.  We first argue that we can
almost surely create such a bisection by populating one of the two
subsets only with the largest component and with monomers and dimers
(components of size 2).
We then use arguments in \cite{istrate-2006-cluster} to
show that having created a bisection in this way, it must be in
the same cluster as {\em all\/} optimal bisections.

As discussed in Section~\ref{sec:width}, in ${\cal G}_{np}$ graphs for
large $n$, a fraction $e^{-\alpha}$ of nodes are monomers.  What about
dimers?  For a pair of nodes to be a dimer, it must contain an edge
(probability $p\approx\alpha/n$), and all other pairs involving one of its two
endpoints ($2n-4$ of these) must not contain an edge (probability
$1-p$): asymptotically, this occurs with probability $\alpha e^{-2\alpha}/n$.
So the expected
number of dimer pairs is $\approx \alpha  e^{-2\alpha} n/2$, and
a.a.s.\ the fraction of nodes in dimers is $\alpha  e^{-2\alpha}$.
Thus, a.a.s.\ the fraction of nodes in monomers or dimers is
$s=e^{-\alpha} (1+\alpha e^{-\alpha})$.

For $\alpha\leq 1$, $s>1/2$.  For $1<\alpha<2\log 2$, $s$ falls below
$1/2$, but one may verify that $g+s>1/2$ where $g$ is the size of the
giant component given in Eq.~(\ref{eq:gc}).  Therefore, as long as
$\alpha<2\log 2$, we a.a.s.\ surely have enough monomers and dimers to
fill up a subset that otherwise contains only the largest component (or
one of them, if not unique).  Call such optimal solutions {\em clean
bisections\/}.

Next, we show that any two distinct clean bisections $A$ and $B$ (and
hence all clean bisections) must be
in the same solution cluster.  If the largest component is unique, $A$
and $B$ can differ at most by the choice of which monomers and
dimers are selected to be in its subset.  If the largest component is
not unique, it cannot be a giant component, so two clean bisections can
then differ at most by: 1) a component consisting of a vanishing
fraction of nodes, and 2) monomers and dimers.  In both cases, we can
take any one of the differing components in $A$ and swap it with monomers and
dimers from the other subset, producing an adjacent clean bisection.
By iterating the process, we will necessarily arrive at $B$ through a
chain of adjacent clean bisections, since we have already shown that
there is a sufficient supply of monomers and dimers for this purpose.

Finally, it is straightforward to see that any optimal solution must be in the
same cluster as some clean bisection.  This holds for almost the same
reason as in the previous paragraph.  If the subset containing the
largest component also contains any components other than monomers and
dimers, each of these asymptotically consists of a vanishing
fraction of nodes.  They may be iteratively swapped with monomers and
dimers to produce a chain
of adjacent solutions, leading to a clean bisection.

Since all clean bisections are in the same cluster, and any
solution is in the same cluster as some clean bisection, all solutions
must lie within a single cluster for $\alpha<2 \log 2$.  This is a
replica symmetric phase.

\subsection{$\alpha\geq 2\log 2$}

When the bisection width is nonzero, the situation is more complicated.  Now
we must consider the different ways in which optimal bisections can cut
the giant component.  We will show that up to some
$\alpha_d>\alpha_c$, optimal bisections avoid a central core of the
giant component and are restricted to a well-defined periphery.  We
then give arguments suggesting that for those graphs, all optimal
bisections are in a single cluster.

\subsubsection{The expander core}

Call the giant component $S$, and imagine that for any $\alpha\geq 2\log
2$, some
identifiable part of it, $X\subseteq S$, is an {\em
expander\/}.  This means that cutting off any piece $X_0$ from $X$ requires
(asymptotically) at least some finite {\em expansion\/}
\begin{equation}
h=\frac{|E(X\setminus X_0,X_0)|}{\min(|X\setminus X_0|,|X_0|)},
\end{equation}
i.e., the number of
cut edges per vertex in the broken piece.  The expander $X$ would then be
precisely the central core that optimal bisections avoid, for
$\epsilon=\alpha-\alpha_c$ less than some constant.  The reason is as
follows.  We have seen in Section \ref{sec:width} that when cutting
trees {\em only\/}, the smallest tree needed is of size
\begin{equation}
t_0> \log(1/\epsilon),
\end{equation}
and so the largest expansion is
\begin{equation}
\label{eq:bound5}
h< \frac{1}{\log(1/\epsilon)}.
\end{equation}
For sufficiently small $\epsilon$, this must fall below the (finite)
value of $h$ for the expander.  Thus, below some finite $\epsilon_0>0$,
trimming the giant component's excess requires fewer cuts if these are
performed exclusively within the mantle than if any part of the cuts are
performed within the core.

\begin{figure}
\begin{center}
\includegraphics[height=3in]{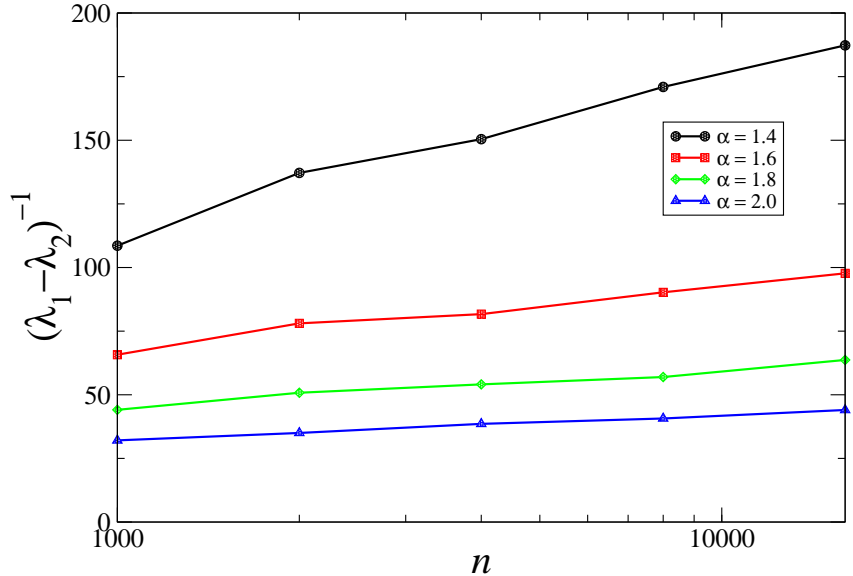}
\end{center}
\vspace{-0.2in}
\caption{Reciprocal of spectral gap.  For mean degree $\alpha=1.4$ through
$\alpha=2.0$, on graphs of size $n=1000$ up to $n=16000$ this quantity
vanishes roughly as $\sim 1/\log n$.\label{fig:gap}}
\end{figure}

One might hope that the 2-core is an expander, in which case greedily
trimming trees would in fact be the optimal way to bisect a graph.  As
it happens, this is almost certain not true.  Consider the 2-core's
{\em spectral gap,\/} the difference $\lambda_1-\lambda_2$ between
the two largest eigenvalues of its connectivity matrix.  Via Cheeger's
inequality~\cite{cheeger-1970-lower}, this yields an upper bound on
expansion.
Figure~\ref{fig:gap} suggests that the spectral gap likely vanishes
(albeit slowly) as
$n\to\infty$, implying that the 2-core's expansion vanishes,
presumably due to the existence of cycles of length $\sim\log n$.  Certain
``bottlenecks'' in the 2-core could then provide more efficient cuts
than trees in the mantle.

\begin{figure}
\begin{center}
\includegraphics[height=3in]{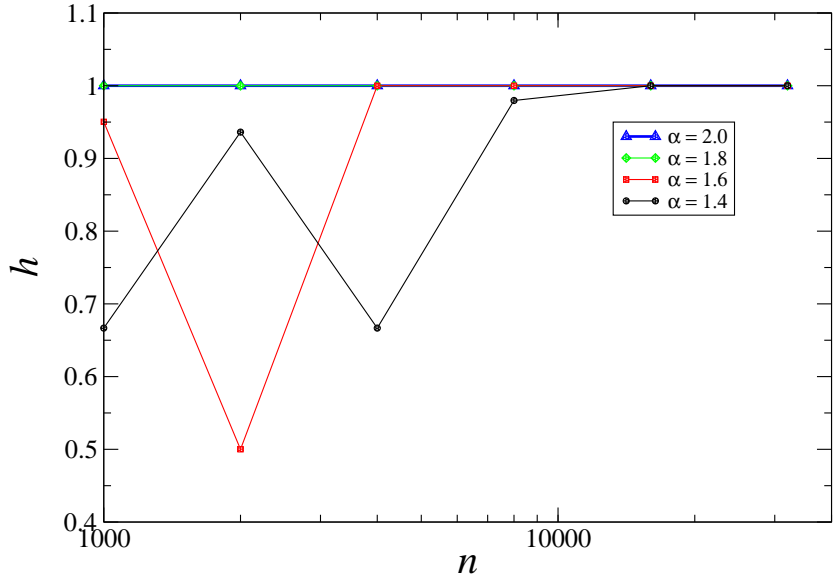}
\end{center}
\vspace{-0.2in}
\caption{Smallest expansion over a random sample of ways of slicing the
2-core into two pieces.  Randomly chosen nodes are sequentially
separated from the 2-core until half of the 2-core has been cut.  For
each value of mean degree $\alpha$, and for each size from $n=1000$ to
$n=32000$, this random sequence is repeated 10 times on each of 100
graphs.  Results show the {\em lowest\/} expansion value found over the
course of all 10 sequences and over all 100 graphs.  For increasing
graph sizes, this value approaches 1.\label{fig:expansion}}
\end{figure}

But some further intuition comes from considering {\em random\/} cuts in
the 2-core.  Figure~\ref{fig:expansion} shows the minimum expansion from
a number of ways of randomly slicing the 2-core into two pieces.  These
numerical results, which we originally reported in
\cite{goncalves-2006-experimental}, suggest that for large $n$ the
minimum expansion never falls below a constant.  This means that any
bottlenecks that do exist in the 2-core are not immediately apparent,
and some significant part of the 2-core could in fact be an expander.

\begin{figure}
\begin{center}
\includegraphics[height=3in]{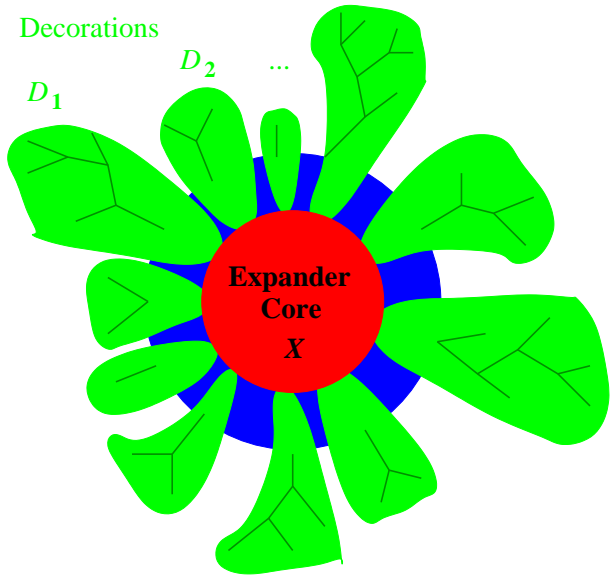}
\end{center}
\vspace{-0.2in}
\caption{A different view of the giant component in
Figure~\ref{fig:strip}, shown here as an identifiable
{\em expander core\/} $X$ within the 2-core.  The trees are now contained within
numerous small {\em decorations\/} $D_i$ that are attached to the expander
core.\label{fig:decorations}}
\end{figure}

Subsequent results by Benjamini et al.~\cite{benjamini-2006-mixing} have
confirmed
that this is indeed the case: for all $\alpha>1$, they give an explicit
procedure for stripping
the giant component down to an expander core.  The giant component is a
``decorated expander,'' with many small ($O(\log n)$ at the largest)
``decorations'' $D_i$ attached to the expander core $X$, as shown
schematically in Figure~\ref{fig:decorations}.
Therefore, by our previous argument, below some
finite $\epsilon_0$, the cuts in all optimal bisections
are restricted to the decorations $D_i$.

\subsubsection{Consequences for solution structure}
\label{sec:structure}

We now use arguments related to the ones in Section 
\ref{sec:alphasmall} to suggest that for $\epsilon<\epsilon_0$
($\alpha<\alpha_d=2\log 2+\epsilon_0$), all optimal solutions are
contained within a single cluster.  We first show how to construct a
{\em neat\/} bisection.  By {\em neat\/}, we mean that one
of the two subsets contains only the expander core, some attached
decorations, and monomers.
We then present arguments suggesting that
this bisection, even if not optimal, is adjacent to a
neat optimal bisection that is in the same cluster as {\em all\/}
optimal bisections, thus replica symmetry holds for all
$\alpha<\alpha_d$.

The construction procedure is as follows.  Consider the giant component
$S$ as being composed of the expander core $X$ and decorations $D_i$ as
in Figure~\ref{fig:decorations}.  These are disjoint subsets, with some
edges connecting the $D_i$ to $X$ but none connecting the $D_i$ to each
other~\cite{benjamini-2006-mixing}.  For each decoration $D_i$, find the
nonempty subset $D_i^1 \subseteq D_i$ that minimizes the expansion
\begin{equation} h_i^1 = \frac{|E(S\setminus D_i^1,D_i^1)|}{|D_i^1|}
\end{equation} resulting from cutting off the piece $D_i^1$.  The value
of $h_i^1$ may be interpreted as a ``cost'' per node that we attribute
to each node in $D_i^1$.  Now repeat this procedure on $D_i\setminus
D_i^1$, resulting in a new piece $D_i^2$, except that this time we find
the $D_i^2$ minimizing the {\em marginal\/} cost of cutting off $D_i^2$
given that $D_i^1$ is already cut off, i.e.,
\begin{equation}
h_i^2 = \frac{|E(S\setminus (D_i^1 \cup D_i^2),D_i^1 \cup D_i^2)| -
|E(S\setminus D_i^1,D_i^1)|}{|D_i^2|}.
\end{equation}
Recursively perform the procedure, letting
\begin{equation}
h_i^j = \frac{|E(S\setminus \bigcup_{l=1}^j D_i^l,\bigcup_{l=1}^j D_i^l)| -
h_i^{j-1} |D_i^{j-1}|}{|D_i^j|},
\end{equation}
until decoration $D_i$ has been
completely partitioned into disjoint subsets $D_i^1,\dots,D_i^m$.  It is
straightforward to confirm that these, by
construction, are now ordered in increasing marginal cost
$h_i^1\leq\dots\leq h_i^m$, where the total cost $|E(S,D_i)|$ of cutting
the entire decoration is simply equal to the sum of the marginal costs
over all nodes,
\begin{equation}
|E(S,D_i)| = \sum_{j=1}^m h_i^j |D_i^j|.
\end{equation}

Note that $h_i^j$ is the {\em minimum possible\/} cost of cutting a node
in $D_i^j$.  Since cuts in optimal solutions are restricted to
decorations, generating a neat optimal bisection requires
an aggregated ranking of all subsets of all decorations, $D_i^j$,
according to their costs $h_i^j$.  Analogously to the greedy algorithm
in Section~\ref{sec:width}, where we stripped trees in increasing order of
expansion, we now strip the pieces of the decorations until the giant
component's excess has been removed.  As with the greedy algorithm, this
can result in overstripping.  There may also be numerous pieces with an
equal ranking, leaving the choice of which ones to pick in order to
overstrip as little as possible.  But due to the maximal size of the
decorations~\cite{benjamini-2006-mixing},
the largest of these pieces is
$O(\log n)$.  So regardless of which we choose, as with tree-stripping,
we can compensate for the imbalance by transferring some of the extensive
supply of monomers.  The result is a neat bisection, albeit not
necessarily an optimal one.

In principle, by perfectly selecting the final pieces of equal rank
and thereby minimizing the cost of overstripping, we can even form any
neat {\em optimal\/} bisection in this way.  We discuss this in the next
section.  But no matter how
we select the final pieces, it seems highly likely that our construction
gives a solution differing from some optimal one by at most the number of
variables contained in a finite number of these pieces, which is $O(\log
n)$.  Furthermore, it is not difficult to see that all neat (not
necessarily optimal) bisections
that use the identical choice of compensating monomers lie within a single
cluster.  It then follows that among these neat bisections, all those
that are optimal are in a single cluster as well.  And {\em every\/}
neat optimal bisection must then be in the same cluster,
since successively swapping monomers gives a chain of adjacent solutions
connecting them.

Finally, for similar reasons as in Section~\ref{sec:width}, any optimal
bisection must be in the same cluster as some neat optimal
bisection.  If the subset containing the expander core also contains
components other than monomers, each one can have size at most $O(\log
n)$ and can be swapped with monomers from the other subset to
produce an adjacent optimal bisection.  Iterating this process leads to
a neat optimal bisection.  Consequently, the argument implies that for
$2\log 2 < \alpha < \alpha_d$, all optimal bisections lie within a
single cluster, and the replica symmetric phase extends up to
$\alpha_d>\alpha_c$.

Above $\alpha_d$, the solution structure is less clear.  When $\alpha$
is large enough, optimal cuts are no longer necessarily restricted to
the decorations: the expander core likely needs to be cut as well.  In
that case, the sizes of the cut pieces could well be extensive.  One
could easily imagine extensive gaps in the solution structure, leading
to multiple clusters and replica symmetry breaking.  Although we have no
direct evidence of it, we speculate that this is the case for $\alpha$
greater than some finite value, suggesting an RSB phase boundary at
or above $\alpha_d$, i.e., above the critical threshold $\alpha_c$.

\section{Discussion and Conclusions}
\label{sec:discussion}

Our study points to an unusual kind of phase structure for random graph
bisection.  The typical scenario, in random combinatorial optimization
problems displaying a sharp threshold in the objective function's
behavior, is that an RSB transition takes place at or below the
main critical threshold: $\alpha_d\leq \alpha_c$.  By contrast,
graph bisection appears to be replica symmetric up to {\em and beyond\/}
the threshold, with $\alpha_d>\alpha_c$.  We note that this lends support
to an analytical bisection width prediction proposed by Liao in
1987~\cite{liao-1987-prl,liao-1988-pra}.  Let us now comment on several
further aspects of $\alpha_d$, as well as on the intriguing algorithmic
consequences of this phase picture.

\begin{figure}
\vspace{-0.35in}
\begin{center}
\includegraphics[width=3in,angle=-90]{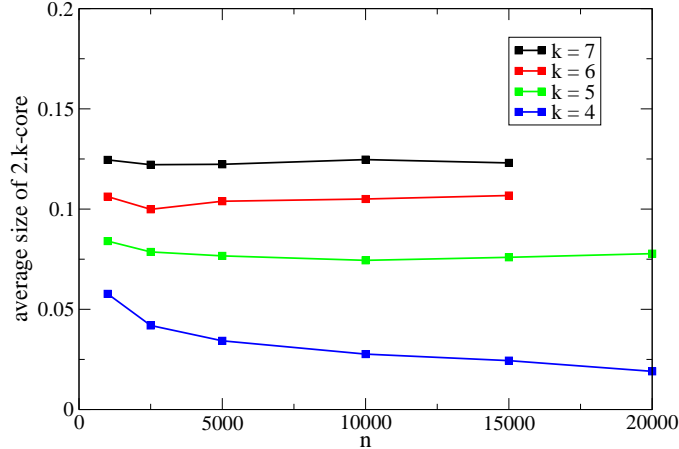}
\end{center}
\vspace{-0.35in}
\caption{Fraction of nodes in the $2.k$-core, $k=4$ through $k=7$, at
$\alpha=1.4$.  Numerical results are averaged over 200 instances, for
increasing values of $n$.\label{fig:2kcore}}
\end{figure}

Although $\alpha_d$ is strictly
greater than $\alpha_c$, it may not be much greater.
As mentioned
at the end of Section~\ref{sec:width}, our bisection width bound
is only valid for $\alpha<1.67$.  That bound
is needed for being able to restrict optimal cuts to the giant
component's decorations.
We do not presently have any other quantitative
bound or estimate on the specific value of $\alpha_d$.  However, some
further qualitative observations come from examining the procedure for
obtaining the expander core.  Roughly speaking, Benjamini et
al.~\cite{benjamini-2006-mixing} strip the giant component down to what
we call a {\em 2.k-core\/}, namely a connected subgraph of the giant
component where all vertices have degree at least 2 (no trees)
{\em and\/} where the largest chain is of length $k$.  The $2.k$-core
is an expander as long as $k$ is a sufficiently large finite constant.
The construction of \cite{benjamini-2006-mixing} does not explicitly
state how large $k$ must be, or what the minimum expansion value $h$ is for the
$2.k$-core at a given mean degree $\alpha$.  But decreasing values of $k$
clearly result in smaller $2.k$-cores (Figure~\ref{fig:2kcore}), and
presumably in larger expansion as more long chains are removed.  This
would lead to a higher $\alpha_d$.
Thus, by picking the smallest possible $k$ (for every $\alpha$) that
still yields an expander core, the resulting expansion gives, via
Eq.~(\ref{eq:bound5}), the largest
possible $\alpha_d$ for which we can be sure that graph bisection is replica
symmetric at all $\alpha<\alpha_d$.

Algorithmically, the striking consequence of our study is that
we may obtain near-optimal bisections and perhaps even optimal ones,
for all $\alpha<\alpha_d$, in
polynomial time.  By near-optimal, we mean that the ratio of its
bisection width to the optimal value approaches 1 asymptotically.
We do this by stripping the giant component down to the expander core
(using any sufficiently large $k$), and then following the construction
in Section~\ref{sec:structure}.  The stripping
process and decoration identification~\cite{benjamini-2006-mixing} takes
polynomial time.  The next subtlety is breaking a decoration
$D_i$ into the correct disjoint pieces $D_i^j$: an exhaustive approach
to finding $D_i^1$ involves comparing all of the $2^{|D_i|}$ ways of
partitioning $D_i$ into two subsets.  But since
no decoration $D_i$ has more than $O(\log n)$ nodes, this can be done in
$n^{O(1)}$ steps, and then all pieces $D_i^j$ can be found and
ranked ($h_i^1\leq\dots\leq h_i^m$) in $n^{O(1)}$ steps.  For the final
pieces of equal rank, even if we simply pick them in arbitrary order we
will overstrip by at most $O(\log n)$.  Thus, in polynomial time we reach
a solution whose (extensive) bisection width is within $O(\log n)$ of the
optimal one.

Of course, the $O(1)$ constant in the exponent may be quite large,
which could make this result largely a theoretical one.  But it is
worth noting that we have some freedom to decrease the size of the
decorations, and thus lower the constant.  If, say,
instead of using the smallest possible $k$ that gives an expander
core, we use the {\em largest\/} possible $k$ that gives expansion
greater than our tree-cutting bound (Eq.~(\ref{eq:bound5})), this will
strip the minimum possible from the giant component.  In this way, we
may be able to minimize the size of the largest decoration.

It would be particularly remarkable if some version of our construction
could give a true optimal bisection, with high probability, in
polynomial time.  For this, we need to pick the final pieces of equal
rank so as to minimize overstripping.  The difficulty is that the
number of such pieces is likely to be infinite: an extensive number
of nodes is contained collectively within the decorations, while there
are at most $O(\log^2 n)$ possible marginal costs ($O(\log n)$ possible
integers for the numerator and $O(\log n)$ possible integers for the
denominator).  On the other hand, it could well be that of the infinite
number of final pieces, a finite fraction of them have any given finite
size $|D_i^j|$.  In that case, a simple greedy procedure will find a
collection of pieces that minimizes overstripping.  Having minimized
overstripping, one still needs to check whether it is possible to lower
the bisection width further by restoring a small fragment of a cut
piece.  If the amount we overstrip is $O(\log n)$, then exhaustively
checking this could require $\sim {n \choose \log n}$ steps.  But more
likely, one can strengthen our analysis to show that the final pieces
are in fact of size $O(1)$, in which case the last operation could also
be accomplished in polynomial time.  Clearly, this intuition is far from
a proof of any sort.  But if indeed the optimal bisection can be found
in polynomial time for $\alpha<\alpha_d$, we would have a very unusual
example of an NP-hard optimization problem whose typical-case complexity
is polynomial at {\em and above\/} the main threshold $\alpha_c$.

We offer two final remarks about other versions of the problem.
First, our approach suggests that the peculiarities of random graph
bisection's phase structure are closely related to the problem's global
constraint --- the fact that a subset cannot be larger than $n/2$ ---
combined with the specific topology of the giant component in ${\cal
G}_{np}$.  This means that the problem of balanced multiway graph
partitioning on ${\cal G}_{np}$ graphs should have very similar
characteristics, and can likely be analyzed through similar means.
Second, the situation is very different if instead of ${\cal G}_{np}$ we
use a graph ensemble that has local geometric structure.  An example is
the random geometric graph ensemble ${\cal
G}_{nr}$~\cite{penrose-2003-book},
where nodes are placed uniformly at random in a unit square and pairs
are connected by an edge if separated by distance $\leq r$.  In this
case, the giant component is very unlikely to have any expander core due
to its large diameter.  Rather than having a tree-like periphery, it
contains numerous local cliques, and minimizing the bisection width
involves locating the bottlenecks in between those cliques.  But at the
same time, ${\cal G}_{nr}$ graphs have useful lattice-like
properties~\cite{goel-2005-monotone}, with the optimal bisection width
scaling~\cite{penrose-2003-book,boettcher-2001-extremal} as
$w\sim\sqrt{n}$, precisely as one finds on a regular 2-dimensional lattice
structure.  We expect that the structure of bisections is closely
related to the geometry of the giant component in bond percolation on a
lattice, which has been studied~\cite{benjamini-2003-mixing}.  Adapting
that analysis could lead to nontrivial bounds on $w$ and rapid
algorithms finding very close approximations to the optimal cut in
${\cal G}_{nr}$.

\section{Acknowledgments}
This work was supported by the U.S.\ Department of Energy
at Los Alamos National Laboratory under contract DE-AC52-06NA25396
through the Laboratory Directed Research and Development Program, a
Marie Curie International Reintegration Grant within the 6th European
Community Framework Programme, a PN II/Parteneriate grant from the
Romanian CNCSIS, and a National Science Foundation grant DMR-0312510.

\bibliographystyle{abbrv}
\bibliography{jmp}
\end{document}